\documentclass[submission,copyright,creativecommons,hyphens]{eptcs}
\providecommand{\event}{VPT 2020} 

\newif\ifVptVer
\VptVertrue
\usepackage{breakurl}             
\usepackage{underscore}           

\usepackage{microtype}

\usepackage{listings}
\usepackage{amssymb}
\usepackage{stmaryrd} 
\usepackage{caption,subcaption}

\usepackage{algpseudocode}      
\usepackage{algorithm}          

\usepackage{graphicx} 

\usepackage[inline]{enumitem}

\usepackage{randtext}
\newcommand{\EmailAddress}[2]{\randomize{#1}@\randomize{#2}}

\ifVptVer
\title{Optimizing Program Size Using Multi-result Supercompilation}
\else
\title{Controlling The Size Of Supercompiled Programs Using Multi-result Supercompilation}
\date{\vspace{-4ex}}
\fi

\ifVptVer
\author{Dimitur Nikolaev Krustev
\institute{IGE+XAO Balkan\\ Sofia, Bulgaria}
\email{\quad \EmailAddress{dkrustev}{ige-xao.com}}
}
\else
\author{Dimitur Krustev\\IGE+XAO Balkan\\ Sofia, Bulgaria \\ \EmailAddress{dkrustev}{ige-xao.com}}
\fi

\ifVptVer
\def\titlerunning{Optimizing Program Size Using MRSC}
\else
\def\titlerunning{Controlling Size Of Supercompiled Programs Using MRSC}
\fi
\def\authorrunning{D.N. Krustev}

\begin{document}
\maketitle

\begin{abstract}
Supercompilation is a powerful program transformation technique with numerous interesting applications.
Existing methods of supercompilation, however, are often very unpredictable with respect to the size
of the resulting programs.
We consider an approach for controlling result size, based on a combination of multi-result
supercompilation and a specific generalization strategy, which avoids code duplication.
The current early experiments with this method show promising results -- we can keep
the size of the result small, while still performing powerful optimizations.
\end{abstract}

\section{Introduction}

Supercompilation was invented by Turchin \cite{TurchinSupercompilerConcept} and has found numerous
applications, such as program optimization\cite{Sorensen1994TurchinSupercompiler,sorm98b,TMR/SCP2014}, 
program analysis, software testing, formal verification \cite{Klyuchnikov2010,Lisitsa2017,MendelGleasonPhD2011}.
It is closely related to partial evaluation \cite{Jones:1993:PEA:153676} and deforestation.
Different extensions of the basic supercompilation approach are also studied, aiming to further increase
its power -- for example, distillation \cite{10.1145/1244381.1244391}, higher-level supercompilation \cite{Klyuchnikov:META2010:HigherLevelScp}, etc.

Supercompilation performs very powerful program transformations by simulating
the actual execution of the input program on a whole set of possible inputs
simultaneously.
The flip side of this power is that the behavior of supercompilation -- 
with respect to both transformation time and result size --
can be very unpredictable.
This makes supercompilation problematic for including as an
optimization step of a standard compiler, for example.
Measures have been proposed to make supercompilation more
well-behaved, both in execution time and result size \cite{bolingbroke2011improving,Jonsson2011Taming},
while still achieving substantial improvements in program run time.
These proposals are all based on a combination of specially crafted and
empirically fine-tuned heuristics.
The main goal of the present study is to experiment with a more
principled approach for finding a better balance between the size and the
run-time performance of programs produced by supercompilation.
This approach is based on a few key ideas:
\begin{itemize}
  \item use multi-result supercompilation\footnote{often abbreviated as \emph{MRSC} from now on} 
    \cite{KlyuchnikovMRSCBranch,Klyuchnikov:META2012:MRSC,Romanenko2014StagedMRSC}
    to systematically explore a large set of different generalizations during 
    the transformation process, 
    leading to different trade-offs between performed optimizations and code explosion;
  \item carefully select a generalization scheme, which can -- if applied systematically -- avoid duplicating
    code during the supercompilation process itself;
  \item re-use ideas from Grechanik et al. \cite{Romanenko2014StagedMRSC} to compactly represent and efficiently
    explore the set of programs resulting from multi-result supercompilation.
\end{itemize}
We outline the main ideas of multi-result supercompilation --
as well as the specific approach to its implementation that we use --
in Sec. \ref{sec:MRSCSummary}. We then describe the main contributions of this study:
\begin{itemize}
  \item We propose a method of adapting some existing techniques (MRSC with
    efficient queries over result sets)
    to solve in a systematic way the problem of code explosion during supercompilation (Sec. \ref{sec:Generalize}).
  \item We define a particular strategy for generalization during MRSC (Sec. \ref{sec:MRDriving}), which:
    \begin{itemize}
      \item avoids any risk of duplicating code during supercompilation (when applied);
      \item avoids unnecessary increase of the search space of possible
        transformed programs, which MRSC must explore.
    \end{itemize}
  \item We analyze the performance of the proposed strategy on several
    simple examples (Sec. \ref{sec:EmpEval}).
\end{itemize}

\section{Summary of Multi-result Supercompilation}\label{sec:MRSCSummary}

Supercompilation is often defined as a transformation of ``configurations'' --
data structures containing information about the set of states of program execution
currently being explored.
The configurations are typically produced by a process called ``driving'',
and organized in ``configuration trees''.
Sometimes the current configuration is similar enough to some previous one,
and we can ``fold'' the former to the latter,
turning the configuration tree into a ``configuration graph''.
Finally, we can generate a new ``residual'' program from the configuration
graph, where folding typically corresponds to calls to functions introduced by new
recursive definitions.
To ensure termination of the supercompilation process, a check -- usually called 
a ``whistle'' -- is systematically performed on the sequence of configurations
being produced.
When this whistle signals a potential risk of non-termination, one possibility
to continue the supercompilation process is to perform a ``generalization'' --
replace the current configuration by another, which avoids the non-termination
risk, possibly by forgetting some information.
The whistle usually marks the current configuration (``lower'' in the tree) as risking
non-termination with respect to some other configuration produced earlier (``higher'' in the tree).
Typical positive supercompilers -- as described, for example, by S{\o}rensen et al. \cite{sorm98b} --
make a choice whether to generalize the ``lower'' or the ``upper'' configuration.
One of the key insights behind multi-result supercompilation is that the place
where the whistle has blown is not always the best place to make a generalization.
The proposed solution is radical: do not generalize when the whistle has blown;
instead, at any driving step check if suitable generalizations exist and
continue driving not only the non-generalized configuration, but also
all generalized ones,
giving rise to multiple alternative results of driving.
This process is illustrated by the example at the end of Sec. \ref{sec:MRDriving}.

Our implementation of multi-result supercompilation mostly follows the same
generic framework used in \cite{Romanenko2014StagedMRSC,krustev2014approach}.
It offers some important simplifications compared to the original work
of Klyuchnikov et al. \cite{KlyuchnikovMRSCBranch,Klyuchnikov:META2012:MRSC}:
\begin{itemize}
  \item It is based on ``big-step'' driving (called so by analogy with big-step
    operational semantics, as driving a configuration produces
    the full configuration subtree corresponding to it).
  \item The whole set of transformed programs is represented compactly
    in a tree-like data structure, which further permits not only
    recovering the full set of configuration graphs, but also
    performing efficiently certain kinds of queries on this set.
\end{itemize}
The compact representation of the set of graphs is shown in Fig. \ref{fig:GraphSet}.
We use direct excerpts of the F\# code of the implementation\footnote{Available at
\url{https://github.com/dkrustev/MRScpOptSize}}, but hopefully
they will be readable by anyone familiar with other functional languages such as OCaml or Haskell.
Another important caveat is that we have implemented MRSC for programs in a specific language,
so some details will only become clear once we introduce this language.
In particular, the configurations (\verb|MConf|) we use are 
pairs of a multi-result driving step output and an expression of the language,
as described in Sec. \ref{sec:MRDriving}.
A language-specific helper function \verb|buildGraph| builds a configuration graph
from a given configuration and subgraphs.
Still, the details of the language are not important for understanding the core
MRSC algorithm; indeed, such details are successfully abstracted away in \cite{Romanenko2014StagedMRSC,krustev2014approach}.
We prefer to show excerpts from our actual implementation for concreteness.
The representation from Fig. \ref{fig:GraphSet} is termed \emph{lazy graph} by Grechanik et al. \cite{Romanenko2014StagedMRSC},
and can be viewed as a domain-specific language (DSL) describing the construction of the
complete set of configuration graphs produced by multi-result supercompilation.
\begin{itemize}
  \item Node \verb|GSNone| is used when the whistle has blown. 
    It represents an empty set of configuration graphs.
  \item Node \verb|GSFold| is used when folding is possible. 
    It gives the relative distance to the upper node to which we fold, 
    plus the renaming (finite mapping of variables to variables), which makes the folded configurations compatible.
    Note that, following Grechanik et al. \cite{Romanenko2014StagedMRSC}, we consider
    only folding to a node on the path from the current node to the root of the graph.
    This choice enables us to keep the representation of the
    set of configuration graphs we produce simpler.
    Also, similar to other positive supercompilers \cite{Sorensen1994TurchinSupercompiler,sorm98b,TMR/SCP2014},
    it is only possible to fold to a node, which is a renaming of the current one.
  \item Node \verb|GSBuild| is the most complicated one, representing a list
    of alternative developments (driving or generalization) of the current configuration.
    Each alternative, in turn, gives rise to a list of new configurations to explore,
    and hence, to a list of nested graph sets.
\end{itemize}

\begin{figure}
\begin{lstlisting}
type MConf = MultiDriveStepResult * Exp

type GraphSet =
  | GSNone
  | GSFold of MConf * int * list<VarName * VarName>
  | GSBuild of MConf * list<list<GraphSet>>

let rec gset2graphs (gs: GraphSet) : seq<MConf * ConfGraph> =
  match gs with
  | GSNone -> Seq.empty
  | GSFold(conf, n, ren) -> Seq.singleton (conf, CGFold(n, ren)) 
  | GSBuild(conf, alts) ->
      let buildGraph' subGraphs = (conf, buildGraph (snd conf) subGraphs)
      let buildAlt alt = 
            Seq.map buildGraph' (Seq.cartesian (Seq.map gset2graphs alt))
      Seq.collect buildAlt alts
\end{lstlisting}
\caption{Representation and Expansion of Graph Sets}
\label{fig:GraphSet}
\end{figure}

The semantics of this DSL is shown in the same Fig. \ref{fig:GraphSet} as
a function \verb|gset2graphs| expanding a \verb|GraphSet| into a sequence 
of configuration graphs.
Note the use of \verb|Seq.cartesian| to compose the subgraphs of 
the graph node of each alternative configuration.

%
%

\begin{algorithm}[ht]
  \caption{Main MRSC Algorithm}
  \label{alg:MRSCAlg}
  \begin{algorithmic}[1]
    \Function{mrScpRec}{$P, l, h, c$}
\ifVptVer    
      \Comment{$P$ -- program (function definitions); $l$ -- nesting level; }
      \State \Comment{$h$ -- history (list of tuples: local/global flag; level; configuration); $c$ -- configuration}
\fi      
      \If{$\exists (\_, l', c') \in h, \rho \textrm{ -- renaming} : c = \textit{rename}(c', \rho)$}
        \State \Return{$\mathtt{GSFold}(c, l - l', \rho)$}
      \Else
        \State $\mathit{rs} \gets \mathtt{multiDriveSteps}(P, c)$
        \If{$\exists (\mathtt{MDSRCases} \_) \in \mathit{rs}$}
          \State $\mathit{hk} \gets \mathtt{HEGlobal}$
          \State $\mathit{relHist} \gets [ (\mathit{hk}', l, c) | \mathit{hk}' = \mathtt{HEGlobal} ]$
        \Else
          \State $\mathit{hk} \gets \mathtt{HELocal}$
          \State $\mathit{relHist} \gets \mathit{takeWhile}(\lambda (\mathit{hk}', l, c) . \mathit{hk}' = \mathtt{HELocal}, h)$
        \EndIf
        \If{$\exists (\_, \_, c') \in \mathit{relHist} : c' \trianglelefteq c$} 
          \Comment{$\trianglelefteq$ denotes \emph{homeomorphic embedding}}
          \State \Return{$\mathtt{GSNone}$}
        \Else
          \State $\mathit{css} \gets \mathit{map}(\mathtt{mdsrSubExps}, \mathit{rs})$
          \State $h' \gets (\mathit{hk}, l, c)::h$
          \State \Return $\mathtt{GSBuild}(c, \mathit{map}(\lambda \mathit{cs} . \mathit{map}(\lambda c . \Call{mrScpRec}{P, l+1, h', c}, \mathit{cs}) , css))$
        \EndIf
      \EndIf
    \EndFunction
    \Function{mrScp}{$P, c$}
      \State \Return $\Call{mrScpRec}{P, 0, [], c}$
    \EndFunction
  \end{algorithmic}
\end{algorithm}

The main MRSC algorithm -- the one that builds the graph set of a given initial
configuration -- is presented with some simplifications as Algorithm \ref{alg:MRSCAlg}.
We can ignore the details about splitting the configuration history into a local
and a global one -- they mostly follow established heuristics as in S{\o}rensen et al. 
\cite{Sorensen1994TurchinSupercompiler,sorm98b}.
The overall approach is simple - if folding is possible, we produce a fold node 
and stop pursuing the current configuration.
Otherwise we check the whistle -- in our case, the same homeomorphic embedding relation,
which is used in other positive supercompilers \cite{sorm98b}.
If it blows, we stop immediately with an empty set of resulting graphs.
When there is neither folding nor a whistle, we continue analyzing the execution of the
current configuration -- based on 2 language-specific functions:
\begin{itemize}
  \item \verb|multiDriveSteps| returns a number of alternatives for the current configuration --
    either a set of new configurations produced by a single step of driving, 
    or by (possibly several different forms of) generalization.
  \item \verb|mdsrSubExps| returns -- for a given alternative produced by the previous function --
    the list of sub-configurations (in our case -- subexpressions) that must be subjected
    to further analysis.
\end{itemize}
The implementation of both functions is described in Sec. \ref{sec:Generalize}.
Once we have this list of lists of sub-configurations, we simply apply the same
algorithm recursively, but with extended history.
Readers familiar with the implementation details of other supercompilers are
invited to compare them to the simplicity of this MRSC approach.

\section{Generalization Approach}\label{sec:Generalize}

\subsection{Programming Language}

The object language we consider is a first-order functional language with ordinary (not pattern-matching) and 
pattern-matching function definitions.
Its syntax is summarized in Fig. \ref{fig:SLLsyntax}.
A very similar language -- with call-by-name semantics -- is often used in many introductions to positive
supercompilation \cite{Sorensen1994TurchinSupercompiler,sorm98b,TMR/SCP2014}.
A notable restriction in our case is that we omit if-expressions and a built-in generic equality.
We use the convention that data constructors always start with an uppercase letter, while
function and variable names start with a lowercase one.
The patterns of any function definition must be exhaustive, not nested, and non-overlapping.
As a technical detail, we do not make a distinction between ordinary and pattern-matching functions
at each call site, as this information is uniquely determined by the function definition itself.

\begin{figure}
\ifVptVer  
\begin{minipage}{0.4\textwidth}
\else
\fi  
\emph{Expressions}
\begin{tabular}[t]{l r l@{\hspace{20pt}} l}
  $e$ & ::= & $x$ & variable     \\
  & $\vert$ & $a(e_1,\ldots,e_n)$ & call 
\end{tabular}
\flushleft{\emph{Call kinds}}
\begin{tabular}[t]{l r l@{\hspace{20pt}} l}
  $a$ & ::= & $C$ & constructor  \\
  & $\vert$ & $f$ & function     
\end{tabular}
\flushleft{\emph{Patterns}}
\begin{tabular}[t]{l r l@{\hspace{20pt}} l}
  $p$ & ::= & $C(x_1, $\ldots$, x_n)$ & 
\end{tabular}
\ifVptVer
\end{minipage}%
\begin{minipage}{0.54\textwidth}
\else
\fi  
\flushleft{\emph{Function definitions}}
\begin{tabular}[t]{l r l@{\hspace{20pt}} l}
  $d$ & ::=     & $f(x_1, \ldots, x_n) = e$ & ordinary function \\
      & $\mid$  & $g(p_1, y_1, \ldots, y_m) = e_1$ & pattern-matching \\
      &         & $\ldots$                         & \hspace{18pt} function \\
      &         & $g(p_n, y_1, \ldots, y_m) = e_n$ & 
\end{tabular}
\flushleft{\emph{Programs}}
\begin{tabular}[t]{l r l@{\hspace{20pt}} l}
  $P$ & ::= & $d_1, \ldots, d_n$ & 
\end{tabular}
\ifVptVer
\end{minipage}
\else
\fi
\caption{Object Language Syntax}
\label{fig:SLLsyntax}
\end{figure}

\subsection{Driving}

Let us recall what driving looks like for this simple language, in the case of positive supercompilation
(which is a simplification of the more general approach pioneered by Turchin \cite{TurchinSupercompilerConcept}).
As a technical device, we define a single step of driving, producing a
result of type \verb|DriveStepResult|, as defined in Fig. \ref{fig:DriveStepResult}

\begin{figure}
\begin{lstlisting}
type DriveStepResult =
  | DSRNone
  | DSRCon of ConName * list<Exp>
  | DSRUnfold of Exp
  | DSRCases of VarName * list<Pattern * Exp>
\end{lstlisting}
\caption{Result of a Single Step of Driving}
\label{fig:DriveStepResult}
\end{figure}

\begin{itemize}
  \item We cannot drive a variable any further: $\mathtt{drive} \llbracket x \rrbracket = \mathtt{DSRNone}$;
  \item Driving a constructor results in a constructor node with all arguments available for
    further driving: $\mathtt{drive} \llbracket C(e_1, \ldots, e_n) \rrbracket = \mathtt{DSRCon}(C, e_1, \ldots, e_n)$;
  \item If we stumble upon a call to an ordinary function, we simply unfold its definition: \\
    $\mathtt{drive} \llbracket f(e_1, \ldots, e_n) \rrbracket$ $=$ $\mathtt{DSRUnfold}(e [ x_1\rightarrow e_1, \ldots, x_n\rightarrow e_n ])$,
    where $f(x_1,$ \ldots$, x_n) = e \in P$ and $e [ x_1\rightarrow e_1, \ldots, x_n\rightarrow e_n ] $
    denotes simultaneous substitution;
  \item The most interesting cases concern a call to a pattern-matching function, as the situation is
    different depending on the kind of the first argument:
    \begin{itemize}
      \item $\mathtt{drive} \llbracket g(C(e'_1, \ldots, e'_m), e_1, \ldots, e_n) \rrbracket =$
        $\mathtt{DSRUnfold}(e [ x_1\rightarrow e'_1, \ldots, x_m\rightarrow e'_m, y_1 \rightarrow e_1, \ldots, y_n \rightarrow e_n ])$
        where $g(C(x_1, \ldots, x_m), y_1, \ldots, y_n) = e \in P$;
      \item $\mathtt{drive} \llbracket g(x, e_1, \ldots, e_n) \rrbracket$ $=$
        $\mathtt{DSRCases}$ $(x,$ $\mathtt{propagate}$ $(x, p_1, (e_1,$ \ldots$, e_n), e'_1),$ \ldots$,$ 
          $\mathtt{propagate}$ $(x, p_m, (e_1,$ \ldots$, e_n), e'_m))$
        where $g(p_1, y_1,$ \ldots$, y_n) = e'_1,$ \ldots$, g(p_m, y_1,$ \ldots$, y_n) = e'_m \in P$
        and $\mathtt{propagate}$ performs positive information propagation by substituting (a suitable renaming of) $p_i$ for $x$ 
        in the corresponding branch $i$;
      \item $\mathtt{drive}$ $\llbracket g(f(e'_1,$ \ldots$, e'_m),$ $e_1,$ \ldots$, e_n) \rrbracket$ $=$
        $\mathtt{dsrMap}$$(\llbracket g(\bullet, e_1,$ \ldots$, e_n) \rrbracket,$ $\mathtt{drive} \llbracket f(e'_1,$ \ldots$, e'_m) \rrbracket)$
        where $\mathtt{dsrMap}$ transforms a driving step result by splicing it in an expression with a hole\footnote{
        We implement expressions with a single hole as functions from expressions to expressions.}.
    \end{itemize}
\end{itemize}
We deliberately omit many low-level details in this description, as they are well-known and can be found
in most introductions to positive supercompilation \cite{Sorensen1994TurchinSupercompiler,sorm98b,TMR/SCP2014}.
Using this definition of driving, plus the usual definitions of folding, whistle, and generalization, 
we can build configuration graphs of the form shown in Fig. \ref{fig:ConfGraph}.
Note that we use the same representation of variables inside object-language programs and inside configuration graphs,
as no confusions arise (assuming suitable measures for avoiding variable capture).

\begin{figure}
\begin{lstlisting}
type ConfGraph =
  | CGLeaf of Exp
  | CGCon of ConName * list<ConfGraph>
  | CGUnfold of ConfGraph
  | CGCases of VarName * list<Pattern * ConfGraph>
  | CGFold of int * list<VarName * VarName>
  | CGLet of list<VarName * ConfGraph> * ConfGraph
\end{lstlisting}
\caption{Representation of a Configuration Graph}
\label{fig:ConfGraph}
\end{figure}

\subsection{Multi-result Driving And Generalization}\label{sec:MRDriving}

As already mentioned, a key difference in multi-result supercompilation is
that driving and generalization are grouped together: a \emph{multi-driving}
step can return not one, but several alternative configurations.
One of them is typically the result of standard driving, but the others
can be different kinds of generalizations.
The choice of generalization strategy depends on the intended use of the
multi-result supercompiler.
In our case, the main goal is to find a program of optimal size among the results.
Previous analyses have shown that one of the main reasons for code size explosion in
supercompilation is the unrestricted duplication of subexpressions during driving.
Of course, sometimes such duplication pays off, as it leads to new opportunities 
for optimization.
But this is not always the case.
These observations lead us to consider two guiding principles that should help us attain our goal:
\begin{itemize}
  \item if standard driving can duplicate existing code, provide also a generalized
    configuration, where no existing (non-trivial) subexpressions are duplicated\footnote{
    We do not attempt to remove already existing code duplication, 
    only to avoid introducing new duplication.};
  \item if there is no risk of duplicating code, avoid any generalization, as it will
    be unlikely to help with the size of the result.
\end{itemize}
To apply these principles, we analyze standard driving, case by case, to see where
we need to avoid code duplication by generalization.
In order to express generalization as a possible result of a driving step,
we extend our representation of a step result - Fig. \ref{fig:MultiDriveStepResult}.
The same figure shows the implementation of \verb|mdsrSubExps| that we have encountered earlier.
The source function \verb|multiDriveSteps| will be denoted $\mathtt{mrdrive}$ for brevity below.

\begin{figure}
\begin{lstlisting}
type MultiDriveStepResult =
  | MDSRLeaf of Exp
  | MDSRCon of ConName * list<Exp>
  | MDSRUnfold of Exp
  | MDSRCases of VarName * list<Pattern * Exp>
  | MDSRLet of list<VarName * Exp> * Exp

let mdsrSubExps (mdsr: MultiDriveStepResult) : list<Exp> =
  match mdsr with
  | MDSRLeaf _ -> []
  | MDSRCon(_, es) -> es
  | MDSRUnfold e -> [e]
  | MDSRCases(_, cases) -> List.map snd cases
  | MDSRLet(binds, e) -> e :: List.map snd binds
\end{lstlisting}
\caption{One Step of Multi-result Driving}
\label{fig:MultiDriveStepResult}
\end{figure}

\begin{itemize}
  \item The variable case is again trivial: \\
    $\mathtt{mrdrive} \llbracket x \rrbracket = [\mathtt{MDSRLeaf}(x)]$;
    (We use $[\ldots; \ldots; \ldots]$ to denote a list of results.)
  \item Driving a constructor does not duplicate code, so we again make no generalization: \\ 
    $\mathtt{mrdrive} \llbracket C(e_1, \ldots, e_n) \rrbracket = [\mathtt{MDSRCon}(C, e_1, \ldots, e_n)]$;
  \item The unfolding of a call to an ordinary function can produce code duplication, if some arguments appear
    multiple times in the body of the definition. 
    We conservatively generalize all arguments of the call\footnote{We leave a more refined generalization treatment for future work.}: \\
    $\mathtt{mrdrive} \llbracket f(e_1, \ldots, e_n) \rrbracket =$ 
      $[\mathtt{MDSRLet}(y_1=e_1,$ \ldots$, y_n=e_n, e [ x_1\rightarrow y_1,$ \ldots$, x_n\rightarrow y_n ]); $ 
      $\mathtt{MDSRUnfold}($ $e [ x_1\rightarrow e_1,$ \ldots$, x_n\rightarrow e_n ])]$,
    where $y_1, \ldots, y_n$ are fresh and $f(x_1, \ldots, x_n) = e \in P$.
    Notice that we shall always place the generalization result before the driving result in the list.
    In this way, when we expand the lazy graph using \verb|gset2graphs|, configuration graphs
    earlier in the resulting sequence will have more generalizations;
  \item The case of a pattern-matching call with a known constructor is completely
    analogous to the previous one: \\
    $\mathtt{mrdrive} \llbracket g(C(e'_1,$ \ldots$, e'_m), e_1,$ \ldots$, e_n) \rrbracket = [$
    $\mathtt{MDSRLet}(u_1=e'_1,$ \ldots$, u_m=e'_m, z_1=e_1,$ \ldots$, z_n=e_n,$ 
      $e [ x_1\rightarrow u_1,$ \ldots$, x_m\rightarrow u_m, y_1 \rightarrow z_1,$ \ldots$, y_n \rightarrow z_n ]);$ 
    $\mathtt{MDSRUnfold}(e [ x_1\rightarrow e'_1,$ \ldots$, x_m\rightarrow e'_m, y_1 \rightarrow e_1,$ \ldots$, y_n \rightarrow e_n ])]$
    where $u_1, \ldots, u_m, z_1, \ldots, z_n$ are fresh and $g(C(x_1, \ldots, x_m), y_1, \ldots, y_n) = e \in P$;
  \item When we pattern-match on a variable, information propagation can introduce some code duplication. 
    The code potentially being duplicated, however, is always of the form $C(x_1, \ldots, x_n)$.
    We have currently decided to accept this limited form of potential duplication, without adding a generalization: \\
    $\mathtt{mrdrive} \llbracket g(x, e_1, \ldots, e_n) \rrbracket =$
    $[$ $\mathtt{MDSRCases}$ $(x,$ $\mathtt{propagate}$ $(x, p_1, (e_1,$ \ldots$, e_n),$ $e'_1),$ \ldots$,$ 
    $\mathtt{propagate}$ $(x, p_m, (e_1,$ \ldots$, e_n), e'_m))]$
    where $g(p_1, y_1,$ \ldots$, y_n) = e'_1,$ \ldots$, g(p_m, y_1,$ \ldots$, y_n) = e'_m \in P$;
  \item The case of matching on a function call is perhaps the least obvious. 
    As during normal driving we reuse the result of driving the nested call, it is not clear in advance what
    it will be.
    So we prefer to be conservative, and add a full generalization of the outer call here: \\
    $\mathtt{mrdrive} \llbracket g(f(e'_1, \ldots, e'_m), e_1, \ldots, e_n) \rrbracket =$
    $[\mathtt{MDSRLet}$ $(x_0=f(e'_1,$ \ldots$, e'_m),$ $x_1=e_1,$ \ldots$, x_n=e_n,$ 
    $g(x_0,$ \ldots$, x_n));$ 
    $\mathtt{mdsrMap}(\llbracket g(\bullet, e_1,$ \ldots$, e_n) \rrbracket,$ $\mathtt{mrdrive} \llbracket f(e'_1,$ \ldots$, e'_m) \rrbracket)]$
    where $x_0, \ldots, x_n$ are, as usual, fresh.
\end{itemize}

%

\begin{figure}
  \centering
  \includegraphics[width=\textwidth]{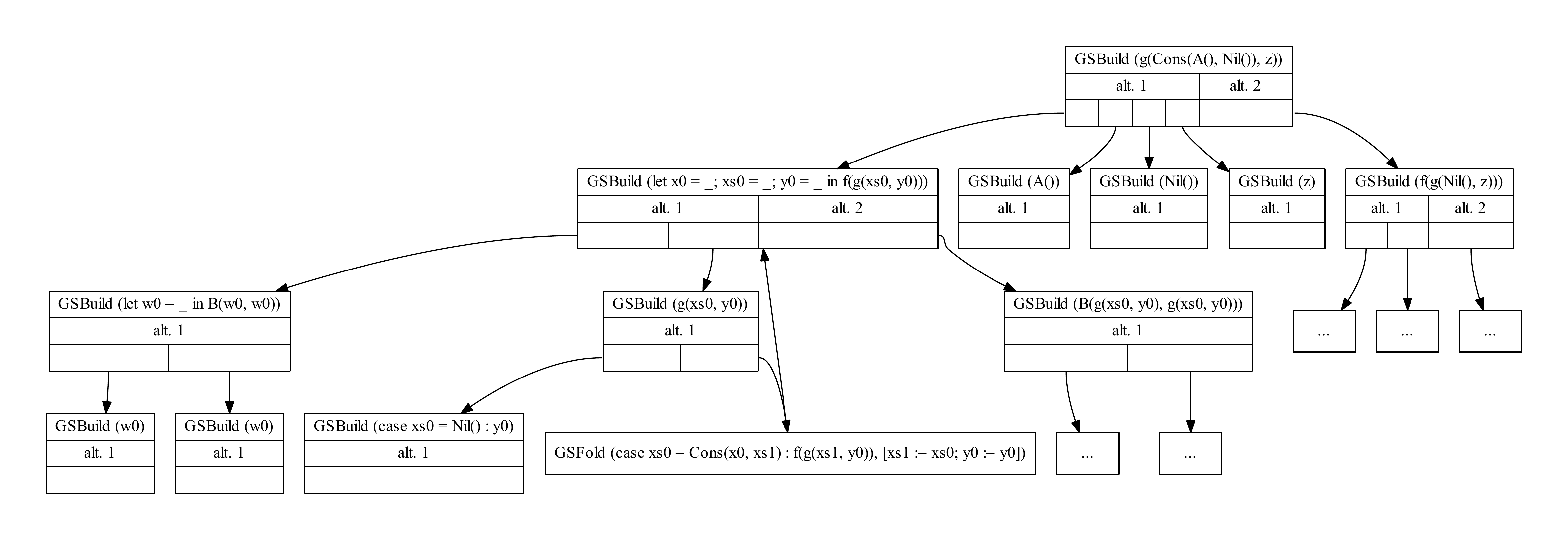}
  \caption{Partial Lazy Graph of ``exp growth'' Small Example}\label{fig:ExpGrowthSmallPartialGraphSet}
\end{figure}

To illustrate the process on a very simple example, consider the example program from 
Fig. \ref{sfig:ExpGrowthPrg}-\subref{sfig:ExpGrowth} (explained in Sec. \ref{sec:EmpEval}), 
but specialized to a smaller input: \verb|g(Cons(A, Nil), z)|.
A part of the resulting lazy graph (omitting subgraphs for alternatives other than the first)
is shown in Fig. \ref{fig:ExpGrowthSmallPartialGraphSet}.
The full lazy graph for the same example is shown in 
\ifVptVer
the extended version of this article \cite{krustev2020controlling}.
\fi
\begin{itemize}
  \item Multi-result driving produces 2 alternatives from the initial expression \verb|g(Cons(A, Nil), z)|:
    \verb|let| \verb|x0 = A;| \verb|xs0 = Nil;| \verb|y0 = z| \verb|in| \verb|f(g(xs0, y0))| and \verb|f(g(Nil, z))|.
    As mentioned, we further consider only the subgraph for the first alternative, which
    is the result of a generalization.
  \item Driving cannot transform any further the subexpressions \verb|A|, \verb|Nil|, and \verb|z|, 
    so they end up as leafs in the lazy graph.
    Driving the subexpression \verb|f(g(xs0, y0))| again produces 2 alternatives:
    \verb|let| \verb|w0 = g(xs0, y0)| \verb|in| \verb|B(w0, w0)| and \verb|B(g(xs0, y0), g(xs0, y0))|.
  \item The first alternative here has 2 subexpressions: 
    \begin{itemize}
      \item \verb|B(w0, w0)|, where driving of its subexpressions in turn leads to 2 leafs, both \verb|w0|;
      \item \verb|g(xs0, y0)|, where driving must perform a case analysis on \verb|x0|, resulting in 2 subgraphs:
      \begin{itemize}
        \item \verb|case xs0 = Nil() : y0|, where driving cannot proceed any further;
        \item \verb|case xs0 = Cons(x0, xs1) : f(g(xs1, y0))| -- this expression is a renaming of
          \verb|f(g(xs0, y0))|, encountered above, so we end up with a folding node.
      \end{itemize}
    \end{itemize}
\end{itemize}

\section{Empirical Evaluation}\label{sec:EmpEval}

We have studied the behavior of the proposed multi-result supercompiler on a few simple
examples, with a focus on the resulting program size.
A straightforward approach, which works for some of the smaller examples,
is to enumerate all resulting configuration graphs and gather statistics from them.
For one example -- the well-known ``KMP test'' -- this approach turned out to be too
time consuming.
So, to make it possible to analyze larger examples, we adapted the approach 
of Grechanik et al. \cite{Romanenko2014StagedMRSC}, which permits to filter
the sequence of configuration graphs produced by MRSC, without explicitly
enumerating them.
In particular, we have implemented functions to extract:
\begin{itemize}
  \item the first configuration graph in the sequence (recall that by the ordering of 
    results during driving, it should contain the most generalizations);
  \item the last one -- with the least number of generalizations;
  \item the graph with the smallest number of nodes;
  \item the graph with the largest number of nodes.
\end{itemize}
\ifVptVer
The implementation of some of these functions is shown in
the extended version of this article \cite{krustev2020controlling}.
\fi
These implementations take polynomial time (and often almost linear time in practice)  
with respect to the size of the lazy graph,
while the number of configuration graphs can be exponential with respect to this size.
As such, they are key to making the proposed approach tractable on larger examples.
By comparing the first to the smallest and the largest graph we can see how 
successful the proposed generalization strategy is in controlling code size.
By comparing the last to the first and the smallest graph, we can see the improvements
that our approach can achieve with respect to standard supercompilation --
as the last configuration graph usually corresponds to the one that a standard positive supercompiler would produce.

After we extract a single configuration graph out of the lazy graph, we can further
produce a new program in the object language from this graph.
This residualization process involves 2 main steps:
  \begin{enumerate*}[label=\itshape\arabic*\upshape)]
    \item Extract a program in a language extended with \verb|case| and \verb|let| expressions from the configuration graph.
      This step also creates recursive function definitions from fold nodes.
    \item Remove \verb|case| and \verb|let| expressions (by a method similar to lambda lifting) to
      obtain a program in the original object language.
  \end{enumerate*}
The residualization process also includes several optimizations:
  \begin{enumerate*}[label=\itshape\alph*\upshape)]
    \item removing ``trivial'' \verb|let| expressions (expressions of the form \verb|let x = y in ...|, or where the variable is
      only used once);
    \item removing duplicated function definitions (a limited form of common subexpression elimination).
  \end{enumerate*}
The results shown in Fig. \ref{fig:DoubleAppResult}-\ref{fig:ExpGrowthOptResult} are all produced by applying
this residualization process on the given configuration graph.
Note, however, that we have decided to compare the sizes of the configuration graphs, and not of the residualized
programs, for a couple of reasons:
\begin{itemize}
  \item The size of the resulting program depends not only on the optimizations performed by supercompilation proper
    (which are reflected directly in the configuration graph), but also on the additional optimizations performed
    during residualization.
    The latter are standard optimizations that can be performed on any program, no matter if it is produced
    by supercompilation or written by hand.
  \item As already mentioned, we can efficiently extract a configuration graph of optimum size from the lazy graph.
    We have no way to do the same with respect to the size of the residual program.
\end{itemize}

\begin{figure*}
\begin{subfigure}[b]{\linewidth}
\begin{lstlisting}[language=Prolog]
append(Nil, ys) = ys;
append(Cons(x, xs), ys) = Cons(x, append(xs, ys));
\end{lstlisting}
\caption{List Append Program}
\label{sfig:AppPrg}
\end{subfigure}

\begin{subfigure}[b]{\linewidth}
\begin{lstlisting}[language=Lisp,keywords={}]
append(append(xs, ys), zs)
\end{lstlisting}
\caption{Double Append}
\label{sfig:DoubleApp}
\end{subfigure}

\begin{subfigure}[b]{\linewidth}
\begin{lstlisting}[language=Lisp,keywords={}]
not(True)  = False;
not(False) = True;

eqBool(True,  b) = b;
eqBool(False, b) = not(b);

match(Nil,         ss, op, os) = True;
match(Cons(p, pp), ss, op, os) = matchCons(ss, p, pp, op, os);

matchCons(Nil,         p, pp, op, os) = False;
matchCons(Cons(s, ss), p, pp, op, os) = matchHdEq(eqBool(p, s), pp, ss, op, os);

matchHdEq(True,  pp, ss, op, os) = match(pp, ss, op, os);
matchHdEq(False, pp, ss, op, os) = next(os, op);

next(Nil,         op) = False;
next(Cons(s, ss), op) = match(op, ss, op, ss);

isSublist(p, s) = match(p, s, p, s);
\end{lstlisting}
\caption{Substring Program}
\label{sfig:SubstrPrg}
\end{subfigure}

\begin{subfigure}[b]{\linewidth}
\begin{lstlisting}[language=Lisp,keywords={}]
isSublist(Cons(True, Cons(True, Cons(False, Nil))), s)
\end{lstlisting}
\caption{``KMP Test''}
\label{sfig:KMP}
\end{subfigure}

\begin{subfigure}[b]{\linewidth}
\begin{lstlisting}[language=Lisp,keywords={}]
eqBool(eqBool(x, y), eqBool(y, x))
\end{lstlisting}
\caption{Bool Equality Symmetry}
\label{sfig:BoolEqSym}
\end{subfigure}

\begin{subfigure}[b]{\linewidth}
\begin{lstlisting}[language=Lisp,keywords={}]
g(Nil,         y) = y;
g(Cons(x, xs), y) = f(g(xs, y));
f(w) = B(w, w);
\end{lstlisting}
\caption{Program Demonstrating Exponential Growth}
\label{sfig:ExpGrowthPrg}
\end{subfigure}

\begin{subfigure}[b]{\linewidth}
\begin{lstlisting}[language=Lisp,keywords={}]
g(Cons(A, Cons(A, Cons(A, Nil))), z)
\end{lstlisting}
\caption{Expression Demonstrating Exponential Growth}
\label{sfig:ExpGrowth}
\end{subfigure}

\caption{Example Programs}\label{fig:Examples}
\end{figure*}

Here we analyze in detail several of the example programs (Fig. \ref{fig:Examples}) showing the most interesting results.
\ifVptVer
The extended version of this article \cite{krustev2020controlling} discusses a few additional examples.
\fi
\begin{itemize}
  \item ``double append'' (Fig. \ref{sfig:AppPrg}-\subref{sfig:DoubleApp}) is traditionally used to 
    demonstrate the power of deforestation and supercompilation.
    It can also be seen as a first step of a proof that list append is associative.
  \item The ``KMP test'' (Fig. \ref{sfig:SubstrPrg}-\subref{sfig:KMP}) is another classical example, 
    which demonstrates the power of supercompilation with respect to deforestation and partial evaluation.
    It involves specializing a sublist predicate to a fixed sublist being
    searched in an unknown list.
  \item ``\verb|eqBool| symmetry'' (Fig. \ref{sfig:SubstrPrg}, \ref{sfig:BoolEqSym}) is intended
    to show that Boolean equality is symmetric.
  \item ``exp growth'' (Fig. \ref{sfig:ExpGrowthPrg}-\subref{sfig:ExpGrowth}) is an example taken 
    from S{\o}rensen's thesis \cite[Example 11.4.1]{Sorensen1994TurchinSupercompiler}, who attributes it to Sestoft.
    It is aimed to demonstrate how classical supercompilation can produce output programs,
    which grow exponentially with respect to the input.
\end{itemize}

The results of the multi-result supercompilation -- using our specific generalization approach --
are summarized in Table \ref{tbl:Stats}.
We give the configuration graph sizes for each of the four types of results (first, last, minimum/maximum size).

\begin{table}
  \centering
  \caption{MRSC Statistics\\(Configuration graph sizes)}\label{tbl:Stats}
  \begin{tabular}{l|r|r|r|r}
    \hline
    Example                 & First & Last & Min. size & Max. size \\ \hline
    double append           &    12 &   10 &        10 &        19 \\
    KMP test                &   203 &   39 &        38 &      1055 \\
    \verb|eqBool| symmetry  &    16 &   17 &        16 &        30 \\
    exp growth              &    15 &   37 &        15 &        57 \\
  \end{tabular}
\end{table}

Several interesting observations arise from analyzing the selected resulting programs themselves:
\begin{itemize}
  \item In 2 cases (``\verb|eqBool| symmetry'' and ``exp growth'') the minimum program coincides with the first (most generalizing) one;
    in 1 case (``double append'') -- with the last (least generalizing) result. 
    In ``KMP test'' the difference between the minimum-size and the last program is minimal.
    This confirms the highly unpredictable impact of driving+generalization on program size.
  \item Note, however, that ``double append'' is a bit of an outlier -- it results in only 3 programs,
    one of which (the first) is isomorphic to the input. The smallest one is the expected
    optimized version, shown in Fig. \ref{fig:DoubleAppResult}.
  \item In all cases, the size of the first result is closer to the minimum than to the maximum size.
    This confirms that our choice of generalization ensures limited growth of the result size.
  \item The smallest/last ``KMP test'' graph produces the expected optimal (as execution time) program,
    as shown in Fig. \ref{fig:KMPResult}.
  \item The last ``\verb|eqBool| symmetry'' graph produces a program, which can indeed serve as evidence of the
    symmetry of Boolean equality -- Fig. \ref{fig:BoolEqSymResult}.
  \item The results of ``exp growth'' are especially interesting in view of our main goal.
    The last result is the same as produced by S{\o}rensen's supercompiler --
    \verb|B(B(B(z, z), B(z, z)), B(B(z, z),| \verb|B(z, z)))| -- clearly suffering from code-size explosion.
    The minimum-size (and also first) program -- shown in Fig. \ref{fig:ExpGrowthMinResult} --
    avoids the pitfall of code-size explosion, thanks to generalization.
    It has, however, also missed some opportunities for static evaluation.
    Interestingly, if we analyze the full set of results, there is another
    graph of size 17 that produces a program, which has eliminated all possible static reductions,
    while avoiding the risk of code explosion -- Fig. \ref{fig:ExpGrowthOptResult}.
    Apparently, if we do not want to miss such results, we need a more refined approach for 
    looking for (close to) minimum-size programs.
    One explanation of this discrepancy is that -- as already explained -- rather than comparing the sizes of 
    the programs produced by residualizing these graphs, we compare configuration graph sizes.
    A possible compromise is to study better size measures for configuration graphs, instead
    of the simple node count we currently use.
    For example, ignoring unfolding nodes when calculating size can give a better idea of the expected size
    of the residualized program, as unfolding nodes are skipped during residualization.
    Another possibility is to find not only (one of) the minimum-size result(s), but the $N$ smallest
    results ($N$ being an input parameter).
\end{itemize}
Based on the last observation above, we have implemented modified queries for finding the graph of minimum (and maximum) size, where
unfolding nodes are not counted -- with very encouraging results:
\begin{itemize}
  \item for ``double append'', ``KMP test'', and ``\verb|eqBool| symmetry'' the modified query returns
    the same optimal programs discussed above, which were also found by the existing queries for 
    minimum or last program;
  \item for ``exp growth'', the modified-minimum query again finds the optimal program -- shown Fig. \ref{fig:ExpGrowthOptResult}
    -- which was missed by all standard queries.
\end{itemize}

\begin{figure}
\begin{lstlisting}
f_0(ys, zs) = f_0_case0(ys, zs);
f_0_case0(Nil(), zs) = zs;
f_0_case0(Cons(x0, xs0), zs) = Cons(x0, f_0(xs0, zs));
f_(xs, ys, zs) = f__case0(xs, ys, zs);
f__case0(Nil(), ys, zs) = f_0(ys, zs);
f__case0(Cons(x00, xs00), ys, zs) = Cons(x00, f_(xs00, ys, zs));
expression: f_(xs, ys, zs)
\end{lstlisting}
\caption{Optimized double-append}
\label{fig:DoubleAppResult}
\end{figure}

\begin{figure}
\begin{lstlisting}
f_1_0_0_0_1_0_0(s0, ss1) = f_1_0_0_0_1_0_0_case0(s0, ss1);
f_1_0_0_0_1_0_0_case0(True(), ss1) = f_1_0_0_0_1_0_0_case1(ss1);
f_1_0_0_0_1_0_0_case0(False(), ss1) = f_0(ss1);
f_1_0_0_0_1_0_0_case1(Nil(), ) = False();
f_1_0_0_0_1_0_0_case1(Cons(s0, ss0), ) = f_1_0_0_0_1_0_0_case2(s0, s0, ss0);
f_1_0_0_0_1_0_0_case2(True(), s0, ss0) = f_1_0_0_0_1_0_0(s0, ss0);
f_1_0_0_0_1_0_0_case2(False(), s0, ss0) = True();
f_0(s) = f_0_case0(s);
f_0_case0(Nil(), ) = False();
f_0_case0(Cons(s0, ss0), ) = f_0_case1(s0, ss0);
f_0_case1(True(), ss0) = f_0_case2(ss0);
f_0_case1(False(), ss0) = f_0(ss0);
f_0_case2(Nil(), ) = False();
f_0_case2(Cons(s0, ss1), ) = f_1_0_0_0_1_0_0(s0, ss1);
expression: f_0(s)
\end{lstlisting}
\caption{Optimized KMP Test Result}
\label{fig:KMPResult}
\end{figure}

\begin{figure}
\begin{lstlisting}
main_case0(True(), y) = main_case2(y);
main_case0(False(), y) = main_case2(y);
main_case2(True(), ) = True();
main_case2(False(), ) = True();
expression: main_case0(x, y)
\end{lstlisting}
\caption{``\texttt{eqBool} symmetry'' Optimal Result}
\label{fig:BoolEqSymResult}
\end{figure}

\begin{figure}
\begin{lstlisting}
f_3(xs0, y0) = f_3_let0(f_3_case0(xs0, y0));
f_3_let0(w0) = B(w0, w0);
f_3_case0(Nil(), y0) = y0;
f_3_case0(Cons(x0, xs1), y0) = f_3(xs1, y0);
expression: f_3(Cons(A(), Cons(A(), Nil())), z))
\end{lstlisting}
\caption{``exp growth'' Minimum-size Result}
\label{fig:ExpGrowthMinResult}
\end{figure}

\begin{figure}
\begin{lstlisting}
main_let1(w0) = B(w0, w0);
expression: main_let1(main_let1(B(z, z)))
\end{lstlisting}
\caption{``exp growth'' Optimal Result}
\label{fig:ExpGrowthOptResult}
\end{figure}

\section{Related Work}

The unpredictability of supercompilation with respect to both performance and
result size is a well-established issue.
Problems with code duplication and result size are discussed by S{\o}rensen \cite{Sorensen1994TurchinSupercompiler}, 
for example.
Few works directly tackle this problem, however.
Bolingbroke et al. \cite{bolingbroke2011improving} study heuristics for improving the general
performance of a specific supercompiler, in order to use it as an automatic phase of an optimizing 
compiler for Haskell.
Some of these heuristics concern avoiding code duplication, and as a consequence they may
lead to improvements in result size, while still producing faster programs.
The key idea is to roll back -- discarding some work done by the supercompiler -- if the heuristics
indicate that this work is not leading to a useful result (a form of generalization).
Speculative execution can also help with code size in some instances.
One problem with this approach is that it is not clear how to generalize it or
apply it to a completely different supercompiler.
The main advantage is that by carefully selecting heuristics, suitable for the specific
supercompiler, the authors report good results on a number of benchmarks.

Jonsson et al. \cite{Jonsson2011Taming} explicitly address both the issue of code explosion and
the related issue of supercompilation time.
The main idea is again to discard the result of supercompiling certain program fragments
if they do not meet certain usefulness criteria (based on the number of reductions performed 
by the supercompiler and the resulting code size).
We can again consider this a form of generalization.
When such generalization happens, however, is based on specific hand-picked heuristics,
apparently based on analyzing the results of different test runs.

Grechanik et al. \cite{Romanenko2014StagedMRSC} propose a generic framework for building ``big-step''
multi-result supercompilers, and a way to efficiently extract results satisfying certain criteria.
Selecting the smallest result is one of the criteria studied.
Optimization of the result size is not a goal of their work, however.
The authors have instantiated the framework on a language simulating counter systems, which
is not Turing-complete, and thus does not demonstrate some of the complications
coming with Turing-complete object languages.
The work of Grechanik et al. \cite{Romanenko2014StagedMRSC} is most closely related to ours: 
we re-use the ideas for implementing
our multi-result supercompiler and for efficiently filtering its results by criteria.
Our main emphasis, however, is on using MRSC together with a generalization strategy, which
is explicitly tailored towards avoiding code duplication and -- consequently -- optimizing result size.
We pay much less attention to supercompilation time, as long as it is not unacceptably big
even for the small examples we want to analyze.

\section{Conclusions and Future Work}

We have presented a study on the feasibility of controlling result size after supercompilation
-- based on using multi-result supercompilation coupled with a specific generalization
strategy avoiding code duplication.
While the idea of multi-result supercompilation is not new, the idea to use it -- combined
with a specific generalization strategy -- for taming code explosion in supercompilation results
appears new.
The current results of the approach -- based on a small set of typical supercompilation examples --
are encouraging:
\begin{itemize}
  \item the smallest configuration graphs we produce do not show exponential growth with respect 
    to the size of the input and typically are much smaller than the largest results;
  \item often the results of small size (though not necessarily the smallest) also feature a
    significant number of optimizations, comparable to what a standard classical supercompiler
    can achieve on the same task.
\end{itemize}

We have already hinted at some areas for potential improvements of the proposed approach:
\begin{itemize}
  \item study less conservative definitions of generalization; for example, avoid generalizing
    expressions, which will not be duplicated (because the corresponding function parameter is
    not referenced multiple times), or expressions, whose duplication is not critical;
  \item study definitions of configuration graph size, which more closely match the expected
    size of the residualized program, to avoid missing interesting results, as was the case
    with ``exp growth''.
\end{itemize}
Clearly the first thing to do, however, is to test the proposed approach on a larger set of
different examples.
Due to the small number of analyzed examples, we consider the current proposal to be work in
progress.
An extended set of tests could give more insight on the strengths and weaknesses of the 
proposed technique, and would likely lead to ideas for further study.

Provided we obtain mostly encouraging results from further testing, the next logical 
step would be to make the approach more practical:
\begin{itemize}
  \item make an implementation covering a larger object language, closer to functional languages
    actually used in practice;
  \item provide a larger set of functions for quickly filtering useful results.
\end{itemize}

From a more theoretical perspective, it would be interesting to try to formulate properties
of generalization, which can give some upper bounds on the code size of MRSC results.
Because of unfolding, which can replace the current configuration with a new one of unrelated size (even
if we avoid code duplication at this point), the task is not trivial.
On the other hand, at least in the case of our simple object language, we have a fixed
list of function definitions, which can give us some bound on the configuration size
after unfolding.

\paragraph{Acknowledgments}
The author would like to thank the four anonymous reviewers for
the helpful suggestions on improving the presentation of
this article.


\bibliographystyle{eptcs}
\bibliography{MRScpOptSize}

\end{document}